\documentclass[11pt,a4paper]{article}
\usepackage{jheppub}
\usepackage[T1]{fontenc}
\usepackage{amsmath}
\usepackage{autobreak}
\usepackage{amssymb}
\usepackage{graphicx}
\usepackage{pdfpages}
\usepackage{slashed}
\usepackage{extarrows}
\usepackage{lmodern}

\usepackage[english]{babel}

\usepackage{hyperref}
\hypersetup{
	linkcolor=blue,
	citecolor=blue,
	urlcolor=blue,
	colorlinks=true
}

\usepackage{microtype}
\newcommand{\NF}{N_F}

\def\ba{\begin{eqnarray}}
\def\ea{\end{eqnarray}}

\newcommand{\myspace}{}

\newcommand{\msbar}{\overline{\mathrm{MS}}}

\newcommand{\Ocal}{\mathcal{O}}

\newcommand{\qb}{\bar{q}}
\newcommand{\qp}{{q^\prime}}
\newcommand{\qbp}{{\bar{q}^\prime}}

\newcommand\numberthis{\addtocounter{equation}{1}\tag{\theequation}}

\makeatletter
\newcommand{\vast}{\bBigg@{3}}
\newcommand{\Vast}{\bBigg@{4}}
\makeatother

\DeclareMathOperator{\rational}{\mbox{Rat}}

\definecolor{red}{HTML}{AE1932}
\definecolor{orange}{HTML}{D16F15}

\allowdisplaybreaks

\title{The Unresolved Behaviour of Polarized Scattering Matrix Elements at NNLO in QCD}
\author{Thomas Gehrmann and} 
\author{Markus L\"ochner}
\affiliation{Physik-Institut, Universit\"at Z\"urich,
  Winterthurerstrasse 190, CH-8057 Z\"urich, Switzerland}
\emailAdd{thomas.gehrmann@uzh.ch}
\emailAdd{markus.loechner@physik.uzh.ch}


\preprint{{\raggedleft
		ZU-TH 71/25
}}

\abstract{
Spin asymmetries in collisions of spin-polarized hadrons probe polarized parton distributions, which encode the spin structure of the colliding hadrons.
To perform precision physics studies with spin asymmetries, higher order QCD corrections to the underlying polarized cross sections are required.
Their numerical implementation relies on the use of an infrared subtraction scheme, which extracts the infrared singular pieces from the real and virtual subprocesses.
We derive the universal behaviour of longitudinally polarized real radiation matrix elements in 
their infrared-singular limits up to next-to-next-to-leading order (NNLO) in QCD, thereby 
 enabling the construction of infrared subtraction schemes for generic polarized cross sections at NNLO.
}

\begin{document} 
\maketitle
\flushbottom

\section{Introduction}
\label{sec:intro}

The structure of the proton is described in terms of parton distribution functions (PDFs), 
which represent the momentum distribution among its partonic constitutents: quarks and gluons.
Unpolarized PDFs are known to an accuracy of a few per cent~\cite{Hou:2019efy,Bailey:2020ooq,NNPDF:2021njg,PDF4LHCWorkingGroup:2022cjn}. The same 
framework is applied to the spin structure of the proton, described in terms of polarized PDFs that 
parametrise the 
 difference of momentum distributions of partons with spin aligned and 
anti-aligned with the proton spin. 
Being based on a sparser data set largely derived from 
fixed-target experiments and the BNL RHIC collider, 
determinations of polarized PDFs~\cite{Borsa:2024mss,Bertone:2024taw,Cruz-Martinez:2025ahf}) remain considerably less accurate than their unpolarized counterparts. 

The extraction of PDFs from a global fit is performed at next-to-leading order (NLO) and next-to-next-to-leading order 
(NNLO) in QCD.
Internally consistent fits require evolution kernels and 
 parton-level subprocess cross sections at the respective order.  
 In the unpolarized case, NNLO corrections~\cite{Heinrich:2020ybq} are available for 
 practically all processes that are included in the global  fits. While the polarized evolution kernels are also known to 
 NNLO~\cite{Moch:2014sna,Blumlein:2021enk,Blumlein:2021ryt}, subprocess corrections to this order 
 are only available 
 for polarized inclusive~\cite{Zijlstra:1993sh,Blumlein:2022gpp} and semi-inclusive~\cite{Bonino:2024wgg,Goyal:2024tmo} 
 deep inelastic scattering (DIS), 
polarized inclusive single-jet production~\cite{Borsa:2020ulb,Borsa:2020yxh}  in DIS,
and for the polarized Drell-Yan process~\cite{Boughezal:2021wjw}.

The upcoming Electron-Ion Collider at BNL~\cite{Accardi:2012qut,Aschenauer:2014cki} will enable a broad range 
of precision measurements in polarized electron-proton collisions, including for example jet or heavy quark production observables.
To fully exploit the upcoming EIC data in the context of NNLO fits of polarized PDFs, the calculation of NNLO corrections for a variety of polarized collider processes is required.
These calculations will typically rely on numerical integration techniques, allowing to mirror the 
final-state definition used in the experimental measurements (fiducial cuts, jet and object reconstruction) in the 
theoretical calculation. The numerical  implementation of higher-order QCD corrections requires 
an infrared subtraction scheme to extract infrared singular contributions from the real and virtual 
subprocesses contributing to an observable under consideration. 

Subtraction schemes rely on the factorization of  any real-radiation squared 
matrix element in its infrared singular limits into the product of a reduced matrix element and 
a universal object called the 
splitting function for collinear limits, or eikonal factor in the soft limit. These are known at NNLO for 
all configurations involving only unpolarized partons: 
splitting functions containing 
up to three collinear particles at tree level~\cite{Campbell:1997hg,Catani:1998nv,Catani:1999ss} and two collinear partons 
at one loop~\cite{Kosower:1999rx,Bern:1999ry},  as well as eikonal 
factors 
for two soft gluons at tree level~\cite{Catani:1999ss} and one soft gluon at one loop~\cite{Catani:2000pi}.  
These results led to the construction of various NNLO subtraction 
schemes~\cite{Binoth:2004jv,Anastasiou:2003gr,Catani:2007vq,Gehrmann-DeRidder:2005btv,Czakon:2010td,Boughezal:2011jf,Boughezal:2015eha,Gaunt:2015pea,Cacciari:2015jma,Caola:2017dug,DelDuca:2016ily,Magnea:2020trj} for 
unpolarized processes, enabling NNLO-accurate predictions for a large range of collider 
processes~\cite{Heinrich:2020ybq}. 

Higher-order corrections to processes with polarized partons involve helicity projectors in terms of 
the $\gamma_5$ matrix and the Levi-Civita tensor, which are defined in a self-consistent manner 
only in four dimensions. Several prescriptions to extend these objects in 
dimensional regularization in $d=4-2\varepsilon$ dimensions have 
been proposed~\cite{tHooft:1972tcz,Breitenlohner:1977hr,Korner:1991sx,Larin:1993tq}, 
thereby introducing an extra layer of complexity into polarized calculations. 
From a practical point of view, the Larin scheme \cite{Larin:1993tq} 
appears to be best-suited for higher order perturbative calculations. 
 It merely increases the algebraic complexity of the calculations, 
 and breaks the axial current Ward identities, which 
 need to be restored by an appropriate renormalization~\cite{Matiounine:1998re, Moch:2014sna, Bonino:2025bqa}.
The renormalization of axial vector vertices and of polarized parton distribution functions takes place when different 
subprocess cross sections are assembled together to yield well-defined and finite predictions for fiducial 
cross sections or inclusive coefficient functions. The  
derivation of bare individual subprocess matrix 
elements is thus performed in a chosen scheme for $\gamma_5$ matrix and the Levi-Civita tensor, whereas
the symmetry-restoring and scheme-dependent renormalization is only applied a posteriori.

In this paper, we derive the unresolved factors that appear in
 NNLO calculations involving 
polarized partons, using the Larin scheme~\cite{Larin:1993tq} throughout. 
Polarization is relevant only if the polarized particle is uniquely identified, either by being in the 
initial state or by fragmenting in the final state.
Both cases exclude configurations with a soft polarized particle, thereby 
 restricting the range of unresolved factors to 
splitting functions with one polarized particle. 
In Section~\ref{sec:method}, we define the kinematics of collinear splittings and describe the extraction 
of splitting functions from colour-ordered matrix elements. Subsequently, we document the 
polarized splitting functions for single-collinear configurations up to one loop in Section~\ref{sec:RV} and 
for triple-collinear configurations at tree level in Section~\ref{sec:RR}. We conclude with an outlook in 
Section~\ref{sec:conc}.

\section{Methodology}
\label{sec:method}
\subsection{Kinematics of collinear splittings}
\label{sec:kin}

The collinear limits of scattering amplitudes and matrix elements  can be studied using a Sudakov parametrization of 
the momenta involved. 
For a set of light-like momenta in the final state, the parametrization reads
\begin{align}
  p_j^\mu = z_j \, p^\mu + k_{T,j}^\mu -\frac{k_{T,j}^2}{2z_j} \frac{n^\mu}{p\cdot n} \,,
  \label{eq:sudakov_n_final}
\end{align}
with $p^2=n^2=0$, $p\cdot n\neq 0$,   $p\cdot k_{T,j} = n\cdot k_{T,j} = 0$ and $k_{T,j}^2\leq 0$ as well as
\begin{equation} 
  \sum_j k_{T,j}^\mu = 0,\qquad  \sum_j z_j = 1.
  \end{equation}
The collinear limit is then attained by all $k_{T,j}^\mu\to 0$, such that  $$ \sum_j p_j^\mu \to p^\mu. $$ 
The 
auxiliary light-like momentum $n^\mu$ fixes the frame of reference. 

In the simple collinear case involving partons $i$ and $j$, \eqref{eq:sudakov_n_final}
implies
\begin{equation}
  s_{ij} = (p_i+p_j)^2 = -\frac{k_T^2}{z(1-z)} \, ,
\end{equation}
such that the collinear system is described by $s_{ij}$ and $z \equiv z_i = 1-z_j$. 
In the triple collinear case of partons $i$, $j$ and $k$, one finds
\begin{align}
  s_{ij} = -\frac{z_i}{z_j}k_{T,j}^2 - \frac{z_j}{z_i}k_{T,i}^2 + k_{T,i}\cdot k_{T,j} \qquad\text{etc.}
\end{align}
for the Mandelstam variables of two collinear partons, which allow to remove scalar products $k_{T,i}\cdot k_{T,j}$ of the transverse momentum vectors of two different particles.
Furthermore, with $k_{T,i}^\mu+k_{T,j}^\mu+k_{T,k}^\mu=0$ one can remove also squares of the transverse momentum vectors by
\begin{align}
k_{T,i}^2 = -k_{T,i}\cdot \left( k_{T,j} + k_{T,k} \right) = z_{i} \, s_{jk} - z_i (1-z_i) s_{ijk} \,.
\end{align}
The triple-collinear system is therefore described by the collinear momentum fractions $z_i$, $z_j$, $z_k=1-z_i-z_j$ and the Mandelstam variables $s_{ij}$, $s_{ik}$, $s_{jk}$.

Collinear emissions off an initial-state parton are described by a parametrization that retains the momentum 
direction of the initial state momentum $p_i^\mu$ for the composite momentum $p^\mu$: 
\begin{equation}
\begin{aligned}
  p_{i}^\mu & = \frac{1}{1-\sum_{r\neq i} x_r}\,p^\mu \,, \\
  p_r^\mu &= -\frac{x_r}{1-\sum_{s \neq i} x_s} p^\mu + k_{T,r}^\mu - \frac{k_{T,r}^2}{2} \frac{1-\sum_{s\neq i} x_s}{x_r} \frac{n^\mu}{p\cdot n} \,,
\end{aligned}
\label{eq:sudakov_if}
\end{equation}
where all momenta $p_{i}^\mu$, $p_r^\mu$ in the collinear cluster and the composite momentum $p^\mu$ 
are still considered considered outgoing. The crossing to the initial state is then obtained by the inversion $(p_i^\mu,p^\mu) \to (-p_i^\mu, -p^\mu)$, also ensuring the space-like nature of $s_{ij}$ and $s_{ik}$.  
Similar collinear momentum parametrizations are given in~\cite{Catani:1996vz,Hoche:2017iem} in the single- and triple-collinear case.

A comparison of~\eqref{eq:sudakov_n_final} and~\eqref{eq:sudakov_if} shows that the kinematics of initial-final splittings 
in the exact collinear limit can be obtained from all-final splittings by substituting
\begin{equation} 
z_i = 1-\sum_{r\neq i} z_r, 
\end{equation}
followed by 
\begin{equation}
z_r = -\frac{x_r}{1-\sum_{r\neq i} x_r}
\end{equation}
for $r\neq i$ in the strict collinear limit. Consequently, initial-final splittings can be derived from final-final splittings and will not be 
considered separately.

\subsection{Angular averaging}
\label{sec:ang}
Upon insertion of the Sudakov parametrization~\eqref{eq:sudakov_n_final}
into a matrix element, all Mandelstam invariants involving the collinear momenta will be expressed 
as linear combinations of scalar products contracting $(p^\mu, n^\mu, k_{T,j}^\mu)$ with the remaining momenta 
$(p_a,\ldots)$ unrelated to the collinear splitting. Upon taking the $k_{T,j}^\mu\to 0$ limit, terms proportional to 
$(k_{T}\cdot p_a)( k_{T}\cdot p_b)/(k_T^2)$ 
will contribute to the dominant scaling behaviour in the collinear limit. They originate from the splitting of a parent gluon 
into collinear partons, and induce angular correlations between the collinear partons and the remaining momenta.

These correlations are only realized for gluon-initiated splittings in processes with sufficiently high multiplicity~\cite{Knowles:1988vs}. They are required 
to render the subtraction of collinear singularities fully local in phase space. For the understanding of the 
unresolved collinear structure, and for the extraction of the infrared singular terms resulting from the 
integration over the associated unresolved phase space, the angular correlations can be safely averaged over. 
Likewise, a subtraction scheme based on angular-averaged subtraction terms can be implemented successfully since the 
average over the unresolved 
 transverse momentum direction is performed as part of the phase space sampling. 

Single collinear limits contain only a  single transverse momentum direction and the average reads as follows:
\begin{align}
& \langle k_T^{\mu} \rangle = 0, \nonumber \\
& \langle k_T^{\mu} k_T^{\nu} \rangle = \frac{k_{T}^2}{d-2} \left( g^{\mu\nu} - \frac{p^{\mu} n^{\nu} + p^\nu n^\mu}{p\cdot n}\right). \label{eq:transverse_momentum_projection_scol}
\end{align}
Notice that only symmetric tensor structures are allowed in this case~\cite{Borsa:2020yxh}.

In triple-collinear limits, a new rank-2 tensor structure $k_{T,i}^{\mu} k_{T,j}^{\nu}$ appears which simultaneously contains two independent transverse momentum vectors.

In order to perform the angular average, it is convenient to parametrize the angular components of the vector $k_{T,j}$ relative to the angular components of $k_{T,i}$.
In four dimensions, the vectors may then be parametrized by
\begin{align}
  k_{T,i} &= \sqrt{-k_{T,i}^2}
  \begin{pmatrix}
    0\\ \cos \varphi \\ \sin \varphi \\0
  \end{pmatrix} \,, &
  k_{T,j} &= \sqrt{-k_{T,j}^2}
  \begin{pmatrix}
    0\\ \cos (\varphi+\Delta \varphi) \\ \sin( \varphi + \Delta \varphi) \\0
  \end{pmatrix} \,,
\end{align}
where $\Delta\varphi$ describes the opening angle between $k_{T,i}$ and $k_{T,j}$.
The azimuthal average is then carried out by integrating over the angle $\varphi$ in four dimensions 
and over $\Omega_{d-2}^{(i)}$ in $d$ dimensions.

The angular average is performed using the general ansatz
\begin{align}
\langle k_{T,i}^{\mu} k_{T,j}^{\nu} \rangle = A \, g^{\mu\nu} + B \, p^\mu p^\nu + C \, p^\mu n^\nu + D \, n^\mu p^\nu + E \, p^\mu p^\nu + F \, \varepsilon^{\mu\nu p n} \,,
  \label{eq:transverse_ansatz_kTi_kTj}
\end{align}
where the last structure is antisymmetric in the indices and is only non-vanishing because of $i\neq j$. It can contribute only to the 
angular average for polarized triple-collinear splitting functions.

For the symmetric part of $k_{T,i}^\mu k_{T,j}^\nu$, the azimuthal average becomes
\begin{align}
\langle k_{T,i}^\mu k_{T,j}^\nu \rangle = \frac{k_{T,i}\cdot k_{T,j}}{d-2} \left( g^{\mu\nu} - \frac{p^\mu n^\nu +p^\nu n^\mu}{p\cdot n} \right) \,.
\end{align}

Unlike the symmetric part of the angular correlations, the antisymmetric part of the angular correlations is problematic in dimensional regularization.
Contracting~\eqref{eq:transverse_ansatz_kTi_kTj} with $\varepsilon_{\mu\nu p n}$ yields 
\begin{align}
\langle \varepsilon_{\alpha\beta\gamma\delta} \, k_{T,i}^\alpha \, k_{T,j}^\beta \, p^\gamma_{\vphantom{T}} \, n^\delta_{\vphantom{T}}
\rangle = F\, \varepsilon_{\mu\nu p n} \, \varepsilon^{\mu\nu p n} \,.
\end{align}
In order to carry out the integral, the Levi-Civita tensor inside the integrand has to be evaluated.
In the Larin scheme, however, this operation is only well-defined if no other  Levi-Civita tensors are 
present in the calculation.  Any further  Levi-Civita tensors, which arise for 
example from helicity projectors of external particles, lead to an ambiguity due to the choice of  
contraction sequence of the indices in $d$ dimensions~\cite{Belusca-Maito:2023wah}.

In practice,  it is sufficient to note that the antisymmetric and the symmetric piece of the tensor are mutually orthogonal.
In four dimensions, this is reflected by the form of the angular correlation between the collinear partons,
\begin{align}
\det(k_{T,i},k_{T,j},p,n) = -\sqrt{k_{T,i}^2 \, k_{T,j}^2}(p\cdot n) \sin(\Delta \varphi) \,,
\end{align}
ensuring independent and non-mixing contributions.

As $k_T^\mu$ is orthogonal to both $p^\mu$ and $n^\mu$, a minimum number of 
independent momentum four-vectors in the reduced matrix element is required to become sensitive on 
the angular-dependent terms.  
The symmetric $k_T$-dependent terms 
require three independent momentum vectors and thus 
start to contribute for reduced four-particle matrix elements, i.e.\ in single collinear 
limits of five-particle processes and in triple collinear limits of six-particle processes. The antisymmetric $k_T$-dependent terms only appear in triple collinear limits and 
require four independent momentum vectors, corresponding to 
reduced five-particle matrix elements. For lower multiplicity processes, the collinear limits immediately yield the 
angular-averaged behaviour of the collinear splitting functions. 

\subsection{Colour ordering}

A convenient representation of the unresolved behaviour of QCD amplitudes can be obtained by representing 
them in a 
colour-ordered basis~\cite{Berends:1987cv,Mangano:1987xk,Mangano:1988kk,Bern:1990ux,DelDuca:1999rs}. 
In this basis, each parton is colour-adjacent only to its direct neighbours in the colour chain, and can only become 
unresolved with respect to these neighbours. Up to NNLO, this colour-adjacency carries over to the level of 
squared amplitudes, where the resulting different colour layers can be identified with squared amplitudes containing 
an increasing number of abelianized gluons (which decouple from all other gluons, and only couple to quarks at the end of the colour chain, i.e.\ behave as photons). 

With all colour and coupling factors removed, the single soft 
limit of 
a colour-ordered $m+1$ parton matrix element 
with gluon $j$ unpolarized then factorizes down to an $m$-parton matrix element and an eikonal factor,
\begin{align}
\Big|(\Delta){M}_{m+1}^{(0)}&(p_1,\dots, p_i, \lambda p_j, p_k, \dots,p_{m+1})\Big|^2 \nonumber \\ &\overset{\lambda \to 0}{\longrightarrow}  \frac{1}{\lambda^2} \frac{2s_{ik}}{s_{ij}s_{jk}} \left|(\Delta){M}_m^{(0)}(p_1,\dots, p_i, p_k, \dots , p_{m+1})\right|^2 + \mathcal{O}\big(\lambda^0\big), \label{eq:single_collinear_fact}
\end{align}
where $s_{i_1\dots i_k} = (p_{i_1}+ \,\cdots \,+ p_{i_k})^2$. 
If gluon $j$ is polarized, the matrix element remains non-singular in the soft limit.

\subsection{Derivation of the universal collinear behaviour}
\label{sec:comp}

In single and triple collinear limits, QCD matrix elements factorize into the product of universal collinear factors (called 
splitting functions) and reduced matrix elements of lower multiplicity, where the collinear particles are combined 
into a single composite momentum. We describe this factorization at tree-level and at one loop in detail in 
Sections~\ref{sec:RV} and~\ref{sec:RR}.  

To compute the splitting functions involving a polarized parton, we closely follow the strategy employed in the 
unpolarized case~\cite{Campbell:1997hg} and extract the collinear behaviour from the simplest matrix elements which contain them.
Single collinear splitting functions are extracted from decay matrix elements of a 
colour-neutral current into three partons (at tree-level and at one loop), while triple collinear splitting functions 
follow from current decay matrix elements into four partons. 

In the unpolarized case~\cite{Campbell:1997hg}, all triple-collinear splitting functions were computed from 
photon-decay matrix elements. Since the photon does not couple directly to gluons, only
quark-induced splittings could be studied from four-parton matrix elements, while 
the extraction of gluon-induced splittings required to consider five-parton matrix elements in their triple-collinear limits. 
To assess gluon-induced splittings in current decay matrix elements with lower multiplicity requires decay 
channels that contain gluons already at Born level in two-particle decays. These can be realized 
using an effective field theory for the decay of a neutralino into a gluino and a gluon~\cite{Haber:1988px} and 
for the Higgs boson decay into gluons~\cite{Wilczek:1977zn,Shifman:1978zn,Inami:1982xt} or the graviton 
decay into gluons~\cite{Han:1998sg}. The 
former two effective theories have been used to derive unpolarized antenna functions~\cite{Gehrmann-DeRidder:2005btv} 
for quark-gluon and 
gluon-gluon configurations to NNLO~\cite{Gehrmann-DeRidder:2005svg,Gehrmann-DeRidder:2005alt} 
and N\textsuperscript{3}LO~\cite{Chen:2023fba,Chen:2023egx}.

The calculation of the current decay matrix elements is performed in dimensional regularization 
in $d=4-2\epsilon$ dimensions.
To extract the polarized splitting functions from 
current decay matrix elements requires the polarization (projection on the difference between spin states) 
of one of the partons and of the current itself.
We employ the Larin prescription~\cite{Larin:1993tq, Zijlstra:1993sh} for the Dirac matrix $\gamma_5$ and the Levi-Civita tensor $\varepsilon^{\mu\nu\rho\sigma}$, which appear in the polarized external projectors.
Their extension to $d$ dimensions in the Larin scheme is, however, non-trivial,
the appropriate projectors were derived in~\cite{Schonwald:2019gmn, Behring:2019tus, Bonino:2025bqa}.

For the decay kinematics 
$$
q^\mu \to p_1^\mu+\ldots + p_n^\mu
$$
with $q^2 \neq 0$ and particle with momentum $p_1^\mu$ polarized, the projectors for 
polarized partons to be applied to the 
squared scattering amplitudes are: 
\begin{eqnarray}
P^{{\rm quark}} &=& \frac{-i}{2p_1\cdot q} \varepsilon_{\mu_1\mu_2\mu_3\mu_4} \, p_{1}^{\mu_3} q^{\mu_4}_{\vphantom{1}} \gamma^{\mu_1}_{\vphantom{1}} \gamma^{\mu_2}_{\vphantom{1}} \slashed{p}{\vphantom{p}}_1\,,\nonumber  \\
P_{\mu_1\mu_2}^{{\rm gluon}} &=&  i\,   \varepsilon_{\mu_1\mu_2\mu_3\mu_4}
\frac{p_1^{\mu_3} n^{\mu_4}_{\vphantom{1}} }{p_1\cdot n}\,.
\end{eqnarray}
with an arbitrary light-like momentum $n$.

The full set of time-like splitting functions  can be obtained from the decay current matrix elements for photons and 
neutralinos. 
The polarized projectors in these cases read: 
\begin{eqnarray}
P_{\mu_1\mu_2}^{{\rm photon}} &=&   i \,\varepsilon_{\mu_1\mu_2\mu_3\mu_4}\,
\frac{q^{\mu_3}_{\vphantom{1}} p_1^{\mu_4} }{p_1\cdot q} \,, \nonumber \\
P^{{\rm neutralino}} &=& \frac{i}{2p_1\cdot q} \varepsilon_{\mu_1\mu_2\mu_3\mu_4}\,  p_{1}^{\mu_3} q^{\mu_4}_{\vphantom{1}} \gamma^{\mu_1}_{\vphantom{1}} \gamma^{\mu_2}_{\vphantom{1}} \slashed{q}\, . 
\end{eqnarray}
We also computed polarized graviton decay matrix elements, using the projectors
documented in~\cite{Moch:2014sna}, which served as cross-check on the polarized splitting functions
obtained from the photon and neutralino decay matrix elements.

\section{Single-collinear behaviour up to one loop}
\label{sec:RV}

Single-collinear limits are present in any current decay matrix element into $n=3$ or more partons. We use the simplest 
(three-parton) decay matrix element to extract the corresponding single-collinear behaviour, which is expressed 
in terms of universal splitting functions. Owing to the simplicity of the three-parton kinematics, angular-dependent 
terms are not resolved, and the resulting splitting functions are angular-averaged. 
 Since the tree-level matrix elements are singular objects in the unresolved regions of the phase space, subleading terms in the dimensional regularization parameter $\varepsilon$ in $d=4-2\varepsilon$ dimensions must be retained in the 
 splitting functions.

\subsection{Tree-level spltting functions}
\label{sec:single_collinear}

\begin{figure}[tb]
	\centering
	\begin{align*}
	\begin{gathered}
	\includegraphics[scale=1]{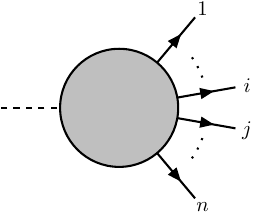}
	\end{gathered}
	\qquad
	\xlongrightarrow{i\parallel j}
	\qquad
	\begin{gathered}
	\includegraphics[scale=1]{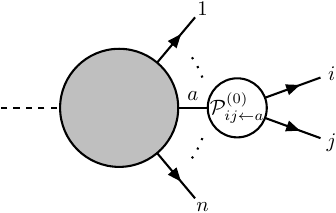}
	\end{gathered}
	\end{align*}
	\caption{Factorization of a tree-level matrix element in the final-final state single-collinear limit $i\parallel j$.}
	\label{fig:single_coll_ff}
\end{figure}

The single-collinear limit between partons $i$, $j$ is taken in the Sudakov parametrization of their 
momenta \eqref{eq:sudakov_n_final}
in the limit where the transverse momentum $k_T$ tends to zero, corresponding to $s_{ij}\to 0$. 
Upon angular average, the matrix element then factorizes as shown in Figure~\ref{fig:single_coll_ff}:
\begin{align}
\left|(\Delta)\mathcal{M}_{m+1}^{(0)}(p_1,\dots,p_i,p_j,\dots p_{m+1})\right|^2 \overset{i\parallel j}{\longrightarrow} \, &  8 \pi \alpha_s \frac{ \mathbf{P}_{(\Delta)ij \leftarrow a}(z)}{s_{ij}} \, |(\Delta)\mathcal{M}_m^{(0)}(p_1,\dots, p, \dots, p_{m+1})|^2 \nonumber \\
& + \mathcal{O}(k_T^0)
\label{eq:singlecol}
\end{align}
where $\mathbf{P}_{ij \leftarrow a}(z)$ is 
the unpolarized splitting function, for $i$ and $j$ both being unpolarized particles, and $\mathbf{P}_{\Delta ij \leftarrow a}(z)$ is the polarized splitting function, for particle  $i$  polarized and particle 
$j$ unpolarized.
In the latter case, the merged particle $a$ with momentum $p$ in the reduced matrix element $\mathcal{M}_m^{0}$ is also polarized. 

At the level of colour-ordered matrix elements, \eqref{eq:singlecol} becomes
\begin{align}
\left|(\Delta){M}_{m+1}^{(0)}(p_1,\dots,p_i,p_j,\dots p_{m+1})\right|^2 \overset{i\parallel j}{\longrightarrow} \, &   \frac{ {P}_{(\Delta)ij \leftarrow a}(z)}{s_{ij}} \, |(\Delta){M}_m^{(0)}(p_1,\dots, p, \dots, p_{m+1})|^2 \nonumber \\
& + \mathcal{O}(k_T^0) \,.
\end{align}

The full set of tree-level single-collinear splitting functions in unpolarized and polarized processes is given by
\begin{alignat}{3}
&\mathbf{P}_{(\Delta) q\, g \leftarrow q} &&= C_F \, && P_{(\Delta) q\, g \leftarrow q}, \nonumber \\
&\mathbf{P}_{(\Delta) g \,q \leftarrow q} &&= C_F \, && P_{(\Delta) g \,q \leftarrow q} , \nonumber \\
&\mathbf{P}_{(\Delta) q \, \bar{q} \leftarrow g} &&= T_R \, && P_{(\Delta) q \, \bar{q} \leftarrow g}, \nonumber \\
&\mathbf{P}_{(\Delta) g \, g \leftarrow g} &&= C_A \, && P_{(\Delta) g \, g \leftarrow g},
\end{alignat}
with the colour-stripped splitting functions, obtained in an exact form in $\varepsilon$,
\begin{align}
P_{qg \leftarrow q}(z) &= \frac{1+z^2}{1-z}-(1-z)\varepsilon, & P_{\Delta q \, g \leftarrow q}(z) &= 
\frac{1+z^2}{1-z}+(1-z)\frac{\varepsilon(3+\varepsilon)}{1-\varepsilon}, \nonumber \\
P_{gq \leftarrow q}(z) &= \frac{1+(1-z)^2}{z} - z \varepsilon , & P_{\Delta g \,q \leftarrow q}(z) &= \frac{2-z(1+\varepsilon)}{1-\varepsilon} , \nonumber \\
P_{q\bar{q} \leftarrow g}(z) &=  1-2 \frac{z(1-z)}{1-\varepsilon}, & P_{\Delta q \, \bar{q} \leftarrow g}(z) &=  1 - 2 \frac{1-z}{1-\varepsilon} , \nonumber \\
P_{gg \leftarrow g}(z) &=  2 z(1-z) + \frac{2}{1-z}+\frac{2}{z}-4 , &  P_{\Delta g \, g \leftarrow g}(z) &= \frac{2}{1-z} -2 + 4 \frac{1- z}{1-\varepsilon}. \label{eq:single_collinear_splitting}
\end{align}

Due to the charge conjugation symmetry of QCD, quarks may also be exchanged by antiquarks.
We notice the genuinely different structure if a polarized particle is involved in the splitting.
The polarized single-collinear tree-level splitting functions were previously computed in the 
't~Hooft-Veltman-Breitenlohner-Maison scheme~\cite{tHooft:1972tcz,Breitenlohner:1977hr} and used to extend 
the dipole subtraction method~\cite{Catani:1996vz} at NLO to 
polarized initial states~\cite{Borsa:2020yxh}. We agree with the 
splitting functions in~\cite{Borsa:2020yxh} through to terms of $\Ocal(\varepsilon^1)$.

The tree-level single-collinear splitting functions $\mathbf{P}_{(\Delta)ij \leftarrow a}(z)$  correspond to the unregulated real radiation part of the corresponding leading-order Altarelli-Parisi evolution kernels $(\Delta) P_{ia}$~\cite{Altarelli:1977zs} and an evanescent $\varepsilon$-dependent term.
The latter depends in the polarized case on the scheme choice for $\gamma_5$.
This becomes apparent for the quark non-singlet splitting function $P_{\Delta q\, g \leftarrow q }(z)$ by considering the finite scheme transformation from the Larin scheme to the $\msbar$ scheme~\cite{Matiounine:1998re}, which is introduced to fix the axial Ward identity and thus restore chriality conservation along the polarized quark line.
This breaking of polarized current conservation originates in the difference 
$$P_{\Delta q \, g \leftarrow q} - P_{qg \leftarrow q}=4(1-z)\varepsilon + \Ocal(\varepsilon^2) \,$$
and is only alleviated by the finite scheme transformation at the level of the finite, mass factorized NLO cross section.

The time-like (final-final) and space-like (initial-final) splitting functions are directly related to each other by the crossing symmetry $z\mapsto z^{-1}$,
\begin{align*}
&P_{qg\leftarrow q}\bigg( \frac{1}{z}\bigg) = -\frac{1}{z} \, P_{qg\leftarrow q}(z) \,, &
&P_{\Delta q \myspace g\leftarrow q}\bigg( \frac{1}{z}\bigg) = -\frac{1}{z} \, P_{ \Delta q \myspace g\leftarrow q}(z) \,, \\
&P_{gq\leftarrow q} \bigg(\frac{1}{z} \bigg) = \frac{1-\varepsilon}{z} \, P_{q\qb\leftarrow g}(z) \,, &
&P_{\Delta g\myspace q\leftarrow q} \bigg(\frac{1}{z} \bigg) = \frac{1}{z} \, P_{\Delta q \myspace \qb\leftarrow g}(z) \,, \\
&P_{q\qb\leftarrow g}\bigg( \frac{1}{z}\bigg) = \frac{1}{(1-\varepsilon) z} \, P_{gq\leftarrow q}(z) \,, &
&P_{\Delta q \myspace \qb\leftarrow g}\bigg( \frac{1}{z}\bigg) = \frac{1}{z} \, P_{\Delta g \myspace q\leftarrow q}(z) \,, \\
&P_{gg\leftarrow g} \bigg(\frac{1}{z} \bigg) = -\frac{1}{z} \, P_{gg\leftarrow g}(z) \,, &
&P_{\Delta g \myspace g\leftarrow g} \bigg(\frac{1}{z} \bigg) = -\frac{1}{z} \, P_{\Delta g \myspace g\leftarrow g}(z) \,. \numberthis
\label{eq:crossing_LO}
\end{align*}
The factors introduced by the crossing relation compensate for signs arising in the crossing and account for changes in the number of degrees of freedom of the crossed particles.
An additional factor $z^{-1}$ is introduced due to rescaling the momentum of the parton which is crossed into the initial state.
The time-like and space-like single-collinear splitting functions thus coincide at tree-level when fixing the external
particle types.

\subsection{One-loop splitting functions}

\begin{figure}[tb]
  \centering
  \resizebox{\textwidth}{!}{
  \begin{minipage}{\textwidth+97.26314pt}
  \begin{align*}
    \begin{gathered}
      \includegraphics[scale=1]{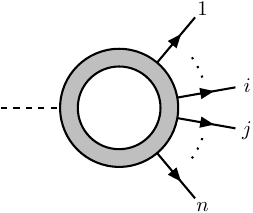}
    \end{gathered}
    \quad
    \xlongrightarrow{i\parallel j}
    \quad
    \begin{gathered}
      \includegraphics[scale=1]{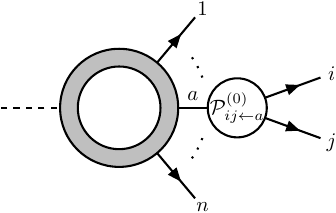}
    \end{gathered}
    \quad
    +
    \quad
    \begin{gathered}
      \includegraphics[scale=1]{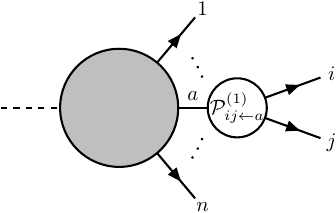}
    \end{gathered}
  \end{align*}
  \end{minipage}
  }
  \caption{Factorization of one-loop matrix element in single-collinear final-final limit $i\parallel j$.}
  \label{fig:single_coll_1L}
\end{figure}

The decomposition of one-loop matrix elements into different colour-ordered contributions proceeds in close analogy 
to the tree-level case.
However, each colour-ordered matrix element (corresponding to the interference of 
tree-level and one-loop scattering amplitudes) now consists of three different colour layers:
\begin{eqnarray*}
& & \bigg[ N_c \left|(\Delta){M}_{m+1}^{(1)}(p_1,\dots,p_i,p_j,\dots p_{m+1})\right|^2 - \frac{1}{N_c} \left|(\Delta)\tilde{M}_{m+1}^{(1)}(p_1,\dots,p_i,p_j,\dots p_{m+1})\right|^2 \nonumber \\ 
&& + \NF \left|(\Delta)\hat{M}_{m+1}^{(1)}(p_1,\dots,p_i,p_j,\dots p_{m+1})\right|^2 \bigg]\,.
\end{eqnarray*}

The singular behaviour of unrenormalized angular-averaged colour-ordered real-virtual one-loop matrix elements is
shown in Figure~\ref{fig:single_coll_1L} and reads as follows~\cite{Kosower:1999rx,Bern:1999ry}:
\begin{eqnarray}
{\left|(\Delta){M}_{m+1}^{(1)}(p_1,\dots,p_i,p_j,\dots p_{m+1})\right|^2 \xlongrightarrow{i\parallel j}
  \frac{ {P}_{(\Delta)ij \leftarrow a}^{(0)}(z)}{s_{ij}} \, |(\Delta){M}_m^{(1)}(p_1,\dots, p, \dots, p_{m+1})|^2  } \nonumber 
  \\ + (-s_{ij})^{-\varepsilon} \,\frac{{P}_{(\Delta)ij \leftarrow a}^{(1)}(z)}{s_{ij}} \, |(\Delta){M}_m^{(0)}(p_1,\dots, p, \dots, p_{m+1})|^2  + \Ocal\big(k_T^0\big) \hspace{1cm}
\end{eqnarray}
for leading colour, likewise
\begin{eqnarray}
{\left|(\Delta)\tilde{M}_{m+1}^{(1)}(p_1,\dots,p_i,p_j,\dots p_{m+1})\right|^2 \xlongrightarrow{i\parallel j}  
  \frac{ {P}_{(\Delta)ij \leftarrow a}^{(0)}(z)}{s_{ij}} \, |(\Delta)\tilde{M}_m^{(1)}(p_1,\dots, p, \dots, p_{m+1})|^2} \nonumber 
\\  +  (-s_{ij})^{-\varepsilon} \,\frac{\tilde{P}_{(\Delta)ij \leftarrow a}^{(1)}(z)}{s_{ij}}  \,  |(\Delta){M}_m^{(0)}(p_1,\dots, p, \dots, p_{m+1})|^2 +  \Ocal\big(k_T^0\big) \hspace{1cm}
\end{eqnarray}
for subleading colour, and
\begin{eqnarray}
{\left|(\Delta)\hat{M}_{m+1}^{(1)}(p_1,\dots,p_i,p_j,\dots p_{m+1})\right|^2 \xlongrightarrow{i\parallel j} 
  \frac{ {P}_{(\Delta)ij \leftarrow a}^{(0)}(z)}{s_{ij}} \, |(\Delta)\hat{M}_m^{(1)}(p_1,\dots, p, \dots, p_{m+1})|^2 } 
  \nonumber \\
+ (-s_{ij})^{-\varepsilon} \,\frac{\hat{P}_{(\Delta)ij \leftarrow a}^{(1)}(z)}{s_{ij}}
 \, |(\Delta){M}_m^{(0)}(p_1,\dots, p, \dots, p_{m+1})|^2 + \Ocal\big(k_T^0\big)\,.
 \hspace{1cm}   
\end{eqnarray}
The unpolarized one-loop splitting functions $P_{ij \leftarrow a}^{(1)}$, $\tilde{P}_{ij \leftarrow a}^{(1)}$
and $\hat{P}_{ij \leftarrow a}^{(1)}$
were derived in~\cite{Kosower:1999rx,Bern:1999ry}. The associated factorization behaviour can be generalized to 
higher loop orders~\cite{Bern:2004cz} and the unpolarized single collinear splitting functions are known 
to two loops~\cite{Bern:2004cz,Badger:2004uk,Duhr:2014nda} and three loops~\cite{Guan:2024hlf}. 

While the tree-level splitting functions are crossed trivially between final-state and initial-state splittings, some care has to be taken in the one-loop splitting functions. The underlying one-loop integrals (bubbles and boxes) must 
be analytically continued to the regions relevant to the splitting, thereby introducing terms dependent on the kinematics.

For definiteness, we first derive the one-loop single-collinear splitting functions $P_{ij \leftarrow a}^{(1)}(z)$  in Euclidean
kinematics, corresponding to an all-final configuration of the momenta with $q^2 < s_{12}, s_{13}, s_{23} < 0$
and the momentum parametrization~\eqref{eq:sudakov_n_final}. 

The $\varepsilon$-exact expressions for the one-loop unpolarized splitting functions~\cite{Kosower:1999rx,Bern:1999ry,Badger:2004uk,Catani:2011st,Braun-White:2023zwd} then read as follows:
\begin{alignat*}{2}
&P_{qg\leftarrow q}^{(1)}(z)&&= c_{\Gamma} \, e^{\varepsilon \gamma_\mathrm{E}} \vast[P_{qg \leftarrow q}^{(0)}(z) \vast(-\frac{(1-z) \, _2F_1\left(1,1+\varepsilon ;2+\varepsilon ;1-\frac{1}{z}\right)}{\varepsilon  (\varepsilon +1) z} \\
&&&\phantom{= c_{\Gamma} \, e^{\varepsilon \gamma_\mathrm{E}} \vast[} -\frac{\Gamma (1-\varepsilon ) \Gamma (1+\varepsilon ) z^{\varepsilon } (1-z)^{-\varepsilon }}{\varepsilon ^2}\vast)+\frac{-\varepsilon +\varepsilon  z+1}{2 (1-2 \varepsilon )}\vast] \,, \numberthis \\
&\tilde{P}_{qg\leftarrow q}^{(1)}(z)&&= c_{\Gamma} \, e^{\varepsilon \gamma_\mathrm{E}} \vast[P_{qg \leftarrow q}^{(0)}(z) \vast(\frac{1}{\varepsilon ^2}-\frac{\, _2F_1\left(1,-\varepsilon ;1-\varepsilon ;1-\frac{1}{z}\right)}{\varepsilon ^2}\vast)-\frac{-\varepsilon +\varepsilon  z+1}{2 (1-2 \varepsilon )}\vast] \,, \numberthis\\
&\hat {P}_{qg\leftarrow q}^{(1)}(z)&&= 0\,, \numberthis \\
\\
&P_{gq\leftarrow q}^{(1)}(z)&&= c_{\Gamma} \, e^{\varepsilon \gamma_\mathrm{E}} \bigg[\frac{1-\varepsilon  z}{2 (1-2 \varepsilon )}-\frac{1}{\varepsilon ^2} \, P_{gq\leftarrow q}^{(0)}(z) \, _2F_1\left(1,-\varepsilon ;1-\varepsilon ;1-\frac{1}{z}\right)\bigg] \,, \numberthis \\
&\tilde{P}_{gq\leftarrow q}^{(1)}(z)&&= c_{\Gamma} \, e^{\varepsilon \gamma_\mathrm{E}} \vast[P_{gq\leftarrow q}^{(0)}(z) \vast(\frac{1}{\varepsilon ^2}-\frac{(1-z) \, _2F_1\left(1,1+\varepsilon ;2+\varepsilon ;1-\frac{1}{z}\right)}{\varepsilon  (1+\varepsilon ) z} \\
&&&\phantom{= c_{\Gamma} \, e^{\varepsilon \gamma_\mathrm{E}} \vast[} -\frac{\Gamma (1-\varepsilon ) \Gamma (1+\varepsilon) z^{\varepsilon } (1-z)^{-\varepsilon }}{\varepsilon ^2}\vast)-\frac{1-\varepsilon  z}{2 (1-2 \varepsilon )}\vast] \,, \numberthis \\
&\hat{P}_{gq\leftarrow q}^{(1)}(z)&&=  0 \,, \numberthis \\
\\
&P_{q\qb\leftarrow g}^{(1)}(z)&&= c_{\Gamma} \, e^{\varepsilon \gamma_\mathrm{E}} \, P_{q\qb\leftarrow g}^{(0)}(z) \vast(\frac{8 \varepsilon ^2-19 \varepsilon +12}{2 (1-2 \varepsilon ) (3-2 \varepsilon ) \varepsilon ^2}-\frac{\, _2F_1\left(1,-\varepsilon ;1-\varepsilon ;1-\frac{1}{z}\right)}{\varepsilon ^2} \\
&&& -\frac{(1-z) \, _2F_1\left(1,1+\varepsilon ;2+\varepsilon ;1-\frac{1}{z}\right)}{\varepsilon  (\varepsilon +1) z}-\frac{\Gamma (1-\varepsilon ) \Gamma (1+\varepsilon ) z^{\varepsilon } (1-z)^{-\varepsilon }}{\varepsilon ^2}\vast) \,, \numberthis \\
&\tilde{P}_{q\qb\leftarrow g}^{(1)}(z)&&= - c_{\Gamma} \, e^{\varepsilon \gamma_\mathrm{E}} \, P_{q\qb\leftarrow g}^{(0)}(z) \, \frac{\left(2 \varepsilon ^2-\varepsilon +2\right) }{2 (1-2 \varepsilon ) \varepsilon ^2} \,, \numberthis \\
&\hat{P}_{q\qb \leftarrow g}^{(1)}(z)&&= - c_{\Gamma} \, e^{\varepsilon \gamma_\mathrm{E}} \,  P_{q\qb\leftarrow g}^{(0)}(z)\, \frac{2 (1-\varepsilon )}{(1-2 \varepsilon ) (3-2 \varepsilon ) \varepsilon } \,, \numberthis \\
\\
&P_{gg\leftarrow g}^{(1)}(z)&&= c_{\Gamma} \, e^{\varepsilon \gamma_\mathrm{E}} \vast[ \frac{2 \varepsilon  z^2-2 \varepsilon  z+1}{(1-2 \varepsilon ) (3-2 \varepsilon ) (1-\varepsilon )} + P_{gg\leftarrow g}^{(0)}(z) \vast(\frac{1}{\varepsilon ^2}-\frac{\, _2F_1\left(1,-\varepsilon ;1-\varepsilon ;1-\frac{1}{z}\right)}{\varepsilon ^2} \\
&&&\phantom{= c_{\Gamma} \, e^{\varepsilon \gamma_\mathrm{E}} \vast[} -\frac{(1-z) \, _2F_1\left(1,1+\varepsilon ;2+\varepsilon ;1-\frac{1}{z}\right)}{\varepsilon  (1+ \varepsilon) z} \\
&&&\phantom{= c_{\Gamma} \, e^{\varepsilon \gamma_\mathrm{E}} \vast[} -\frac{\Gamma (1-\varepsilon ) \Gamma (1+\varepsilon ) z^{\varepsilon } (1-z)^{-\varepsilon }}{\varepsilon ^2}\vast) \vast] \,, \numberthis \\
&\tilde{P}_{gg\leftarrow g}^{(1)}(z)&&=  0  \,, \numberthis \\
&\hat{P}_{gg\leftarrow g}^{(1)}(z)&&= -c_{\Gamma} \, e^{\varepsilon \gamma_\mathrm{E}} \frac{2 \varepsilon  z^2-2 \varepsilon  z+1}{(1-2 \varepsilon ) (3-2 \varepsilon ) (1-\varepsilon )^2} \,, \numberthis \\
\end{alignat*}
Here we defined
\begin{align}
 & c_\Gamma \equiv \frac{\Gamma(1-\varepsilon)^2 \, \Gamma(1+\varepsilon)}{\Gamma(1-2\varepsilon)}
\end{align}
and $\gamma_\mathrm{E}\simeq 0.5772$ denotes the Euler-Mascheroni constant.

The newly derived one-loop polarized splitting functions, obtained in the Larin scheme, take the following 
$\varepsilon$-exact form: 
\begin{alignat*}{2}
&P_{\Delta q \myspace g\leftarrow q}^{(1)}(z)&&= c_{\Gamma} \, e^{\varepsilon \gamma_\mathrm{E}} \vast[P_{\Delta q \myspace g \leftarrow q}^{(0)}(z) \vast(-\frac{(1-z) \, _2F_1\left(1,1+\varepsilon ;2+ \varepsilon ;1-\frac{1}{z}\right)}{\varepsilon  (1+\varepsilon) z} \,, \numberthis \\
&&&\phantom{= c_{\Gamma} \, e^{\varepsilon \gamma_\mathrm{E}} \vast[} -\frac{\Gamma (1-\varepsilon ) \Gamma (1+\varepsilon) z^{\varepsilon } (1-z)^{-\varepsilon }}{\varepsilon ^2}\vast)+\frac{\varepsilon ^2+2 \varepsilon -z \varepsilon ^2 -3 \varepsilon  z+1}{2 (1-2 \varepsilon ) (1-\varepsilon )}\vast] \,, \numberthis \\
&\tilde{P}_{\Delta q \myspace g\leftarrow q}^{(1)}(z)&&= c_{\Gamma} \, e^{\varepsilon \gamma_\mathrm{E}} \vast[P_{\Delta q \myspace g \leftarrow q}^{(0)}(z) \left(\frac{1}{\varepsilon ^2}-\frac{\, _2F_1\left(1,-\varepsilon ;1-\varepsilon ;1-\frac{1}{z}\right)}{\varepsilon ^2}\right) \\
&&&\phantom{= c_{\Gamma} \, e^{\varepsilon \gamma_\mathrm{E}} \vast[} -\frac{\varepsilon ^2+2 \varepsilon -\varepsilon ^2 z -3 \varepsilon  z+1}{2 (1-2 \varepsilon ) (1-\varepsilon )}\vast] \,, \numberthis \\
&\hat{P}_{\Delta q \myspace g\leftarrow q}^{(1)}(z)&&= 0 \,, \numberthis \\
\\
&P_{\Delta g \myspace q\leftarrow q}^{(1)}(z)&&= c_{\Gamma} \, e^{\varepsilon \gamma_\mathrm{E}} \bigg[\frac{1-\varepsilon  z}{2 (1-2 \varepsilon ) (1-\varepsilon )}-\frac{1}{\varepsilon ^2} \, P_{\Delta g \myspace q\leftarrow q}^{(0)}(z) \, _2F_1\left(1,-\varepsilon ;1-\varepsilon ;1-\frac{1}{z}\right) \bigg] \,, \numberthis \\
&\tilde{P}_{\Delta g \myspace q\leftarrow q}^{(1)}(z)&&= c_{\Gamma} \, e^{\varepsilon \gamma_\mathrm{E}} \vast[P_{\Delta g \myspace q\leftarrow q}^{(0)}(z) \vast(\frac{1}{\varepsilon ^2}-\frac{(1-z) \, _2F_1\left(1,1+\varepsilon ;2+\varepsilon ;1-\frac{1}{z}\right)}{\varepsilon  (1+\varepsilon) z} \\
&&&\phantom{= c_{\Gamma} \, e^{\varepsilon \gamma_\mathrm{E}} \vast[} -\frac{\Gamma (1-\varepsilon ) \Gamma (1+\varepsilon) z^{\varepsilon } (1-z)^{-\varepsilon }}{\varepsilon ^2}\vast)-\frac{1-\varepsilon  z}{2 (1-2 \varepsilon ) (1-\varepsilon )}\vast] \,, \numberthis \\
&\hat {P}_{\Delta g \myspace q\leftarrow q}^{(1)}(z)&&= 0\,, \numberthis \\
&P_{\Delta q \myspace \qb\leftarrow g}^{(1)}(z)&&= c_{\Gamma} \, e^{\varepsilon \gamma_\mathrm{E}} \, P_{\Delta q \myspace \qb\leftarrow g}^{(0)}(z) \vast( \frac{8 \varepsilon ^2-19 \varepsilon +12}{2 (1-2 \varepsilon ) (3-2 \varepsilon ) \varepsilon ^2}-\frac{\, _2F_1\left(1,-\varepsilon ;1-\varepsilon ;1-\frac{1}{z}\right)}{\varepsilon ^2} \\
&&&  -\frac{(1-z) \, _2F_1\left(1, 1+\varepsilon ; 2+\varepsilon;1-\frac{1}{z}\right)}{\varepsilon  (1+\varepsilon) z} -\frac{\Gamma (1-\varepsilon ) \Gamma (1+\varepsilon) z^{\varepsilon } (1-z)^{-\varepsilon }}{\varepsilon ^2}\vast) \,, \numberthis \\
&\tilde{P}_{\Delta q \myspace \qb\leftarrow g}^{(1)}(z)&&= - c_{\Gamma} \, e^{\varepsilon \gamma_\mathrm{E}} \, P_{\Delta q \myspace \qb\leftarrow g}^{(0)}(z) \, \frac{2 \varepsilon ^2-\varepsilon +2}{2 (1-2 \varepsilon ) \varepsilon ^2} \,, \numberthis \\
&\hat{P}_{\Delta q \myspace \qb \leftarrow g}^{(1)}(z)&&= -c_{\Gamma} \, e^{\varepsilon \gamma_\mathrm{E}} \,  P_{\Delta q \myspace \qb\leftarrow g}^{(0)}(z) \, \frac{2 (1-\varepsilon )}{(1-2 \varepsilon ) (3-2 \varepsilon ) \varepsilon } \,, \numberthis \\
\\
&P_{\Delta g \myspace g\leftarrow g}^{(1)}(z)&&= c_{\Gamma} \, e^{\varepsilon \gamma_\mathrm{E}} \vast[\frac{1}{(1-2 \varepsilon ) (3-2 \varepsilon ) (1-\varepsilon )} + P_{\Delta g \myspace g\leftarrow g}^{(0)}(z) \vast(\frac{1}{\varepsilon ^2} \\
&&&\phantom{=c_{\Gamma} \, e^{\varepsilon \gamma_\mathrm{E}} \vast[} -\frac{\, _2F_1\left(1,-\varepsilon ;1-\varepsilon ;1-\frac{1}{z}\right)}{\varepsilon ^2} -\frac{(1-z) \, _2F_1\left(1,1+\varepsilon; 2+\varepsilon ;1-\frac{1}{z}\right)}{\varepsilon  (1+\varepsilon) z} \\
&&&\phantom{=c_{\Gamma} \, e^{\varepsilon \gamma_\mathrm{E}} \vast[} -\frac{\Gamma (1-\varepsilon ) \Gamma (1+\varepsilon) z^{\varepsilon } (1-z)^{-\varepsilon }}{\varepsilon ^2}\vast)\vast] \,, \numberthis \\
&\tilde{P}_{\Delta g \myspace g\leftarrow g}^{(1)}&&= 0\,, \numberthis \\
&\hat{P}_{\Delta g \myspace g\leftarrow g}^{(1)}&&= -c_{\Gamma} \, e^{\varepsilon \gamma_\mathrm{E}}\frac{1}{(1-2 \varepsilon ) (3-2 \varepsilon ) (1-\varepsilon )^2} \,. \numberthis \\
\end{alignat*}
We recall that the unpolarized and polarized one-loop splitting functions stated above 
correspond to the unresolved limits of unrenormalized one-loop matrix elements, thereby yielding the unrenormalized
splitting functions. 

Separating the $1/\varepsilon^2$ pole, which is proportional to the corresponding tree-level splitting functions, from the remaining rational terms, we observe the following relations between the rational parts in our results, 
in line with the universal structure of the one-loop splitting functions observed in~\cite{Catani:2011st}:
\begin{align}
\rational P_{qg \leftarrow q}^{(1)}(z) &= -\rational \tilde{P}_{qg \leftarrow q}^{(1)}(z) \,, &
\rational P_{\Delta q \myspace g \leftarrow q}^{(1)}(z) &= -\rational \tilde{P}_{\Delta q \myspace g \leftarrow q}^{(1)}(z) \,, 
\nonumber \\
\rational P_{gq \leftarrow q}^{(1)}(z) &= -\rational \tilde{P}_{gq \leftarrow q}^{(1)}(z) \,, &
\rational P_{\Delta g \myspace q \leftarrow q}^{(1)}(z) &= -\rational \tilde{P}_{\Delta g \myspace q \leftarrow q}^{(1)}(z) \,, 
\nonumber \\
\rational P_{gg \leftarrow g}^{(1)}(z) &= -\rational \hat{P}_{gg \leftarrow g}^{(1)}(z) \,, &
\rational P_{\Delta g \myspace g \leftarrow g}^{(1)}(z) &= -\rational \hat{P}_{\Delta g \myspace g \leftarrow g}^{(1)}(z) \,.
\end{align}
We furthermore observe that the splitting functions $P_{q\qb \leftarrow g}^{(1)}(z)$ and $P_{\Delta q \, \qb \leftarrow g}^{(1)}(z)$ are rational functions and coincide with each other up to the multiplicative tree-level splitting function.

Also the subleading colour splitting functions $\tilde{P}_{q\qb \leftarrow g}^{(1)}(z)$ and $\tilde{P}_{\Delta q \, \qb \leftarrow g}^{(1)}(z)$ coincide with each other up to the tree-level splitting function.
They are furthermore obtained by multiplying the leading order splitting function with bare one-loop quark form factors~\cite{Altarelli:1979ub}.
This is explained by the absence of non-abelian couplings in the subleading colour contribution.
As a consequence, only the diagram containing the one-loop vector vertex correction yields a non-vanishing contribution to the one-loop amplitude constituting $\tilde{P}_{(\Delta) q \, \qb \leftarrow g}^{(1)}(z)$.

The splitting functions derived above in Euclidian kinematics form the starting point for their analytic continuation 
to Minkowskian kinematics in final-state (time-like $q^2$ and $0\leq z \leq 1$) and initial-state (space-like $q^2$ and 
$z\geq 1$) 
splittings. 
We start by noting that the hypergeometric functions with argument $1-1/z$ remain on their single-valued principal branch $(-\infty,+1]$ for all $z\geq 0$, thus requiring no 
analytic continuation.
In the final-state splitting, only the overall factor $(-s_{ij})^{-\varepsilon}$ is affected by the analytic continuation, which is achieved by replacing $ (-s_{ij})^{-\varepsilon} \to \cos(\epsilon\pi) \, (|s_{ij}|)^{-\varepsilon}$.
In the initial-state splitting, we have $s_{ij}<0$ and $z>1$.
This leaves 
$(-s_{ij})^{-\varepsilon}=(|s_{ij}|)^{-\varepsilon}$ unchanged, but requires 
the analytic continuation of all terms 
proportional to $(1-z)^{-\varepsilon}$ in the splitting functions.  
This analytic continuation amounts to replacing $(1-z)^{-\varepsilon} \to \cos(\varepsilon\pi) \, (z-1)^{-\varepsilon}$. 

With these prescriptions, the space-like and time-like splitting functions can be related to each other by the relation $z \mapsto z^{-1}$, which requires an analytic continuation in $(1-z)$ as well as keeping proper account of the overall factor $(-s_{ij})^{-\varepsilon}$.
For a given set of external partons, the crossing relations are identical to the tree-level case~\eqref{eq:crossing_LO}.

\section{Triple-collinear behaviour at tree level}
\label{sec:RR}

\begin{figure}[tb]
	\centering
	\begin{align*}
	\begin{gathered}
	\includegraphics[scale=1]{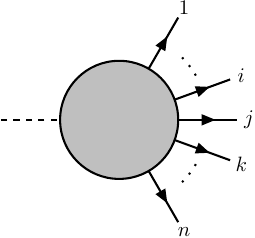}
	\end{gathered}
	\qquad
	\xlongrightarrow{i\parallel j \parallel k}
	\qquad
	\begin{gathered}
	\includegraphics[scale=1]{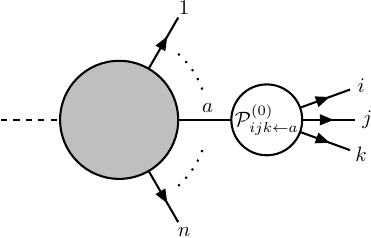}
	\end{gathered}
	\end{align*}
	\caption{Factorization of tree-level matrix element in final state triple-collinear limit $i\parallel j \parallel k$.}
	\label{fig:triple_coll_ff}
\end{figure}

The triple-collinear limit between partons $i$, $j$ and $k$ is obtained by taking the limit of vanishing transverse momenta
$k_{T,n}$ ($n=(i,j,k)$) in \eqref{eq:sudakov_n_final}, corresponding to $s_{ij}+s_{ik}+s_{jk}= s_{ijk} \to 0$.
The triple collinear splitting function is then the residue of the double pole in $s_{ijk}$ and the 
 matrix element factorizes as displayed in Figure~\ref{fig:triple_coll_ff} into
\begin{eqnarray}
\lefteqn{\left|(\Delta)\mathcal{M}_{m+2}^{(0)}(p_1,\dots,p_i,p_j,p_k,\dots p_{m+2})\right|^2 \overset{i\parallel j\parallel k}{\longrightarrow} } \nonumber \\ 
 & &(8\pi \alpha_s)^2 \frac{1}{s_{ijk}^2} \mathbf{P}_{(\Delta)ijk \leftarrow a}\bigg( \frac{s_{ij}}{s_{ijk}}, \frac{s_{ik}}{s_{ijk}}, \frac{s_{jk}}{s_{ijk}}, z_i, z_j, z_k \bigg)  \, |(\Delta)\mathcal{M}_m^{(0)}(p_1,\dots, p, \dots, p_{m+2})|^2 
 \nonumber \\  &&+ \mathcal{O}(k_{T,n}^{-2})\,,
\label{eq:triplecol}
\end{eqnarray}
where $\mathbf{P}_{ijk \leftarrow a}(z)$ is 
the unpolarized splitting function, for $(i,j,k)$ all unpolarized particles, and $\mathbf{P}_{\Delta ijk \leftarrow a}(z)$ is the polarized splitting function, for particle  $i$  polarized and particles 
$(j,k)$ unpolarized. In the latter case, the merged particle $a$ with momentum $p$ in the reduced matrix element $\mathcal{M}_m^{0}$ is also polarized.

In terms of colour-ordered matrix elements, the triple collinear limit yields:
\begin{eqnarray}
\lefteqn{\left|(\Delta){M}_{m+2}^{(0)}(p_1,\dots,p_i,p_j,p_k,\dots p_{m+2})\right|^2 \overset{i\parallel j\parallel k}{\longrightarrow} } \nonumber \\ 
 & &  \frac{1}{s_{ijk}^2} {P}_{(\Delta)ijk \leftarrow a}\bigg( \frac{s_{ij}}{s_{ijk}}, \frac{s_{ik}}{s_{ijk}}, \frac{s_{jk}}{s_{ijk}}, z_i, z_j, z_k \bigg)  \, |(\Delta){M}_m^{(0)}(p_1,\dots, p, \dots, p_{m+2})|^2 
 \nonumber \\ && + \mathcal{O}(k_{T,n}^{-2})\,.
\label{eq:triplecol_ord}
\end{eqnarray}
In taking the triple collinear limits in \eqref{eq:triplecol} and \eqref{eq:triplecol_ord}, angular terms have been
averaged over as outlined in Section~\ref{sec:ang}. 
The subleading terms of $\mathcal{O}(k_{T,n}^{-2})$ in both equations
 are not sufficiently singular to yield infrared-divergent contributions 
 when being integrated over the phase space associated to triple collinear configurations and can therefore be discarded.

According to the particle content and the color ordering, the tree-level unpolarized
splittings give rise to the seven independent triple-collinear splitting functions derived in~\cite{Campbell:1997hg},
\begin{align}
  P_{qgg  \leftarrow q}^{(0)},\,
  P_{q\gamma \gamma  \leftarrow q}^{(0)},\,
  P_{q\bar{q}^\prime q^\prime  \leftarrow q}^{(0)},\,
  P_{q\bar{q}q  \leftarrow q}^{(0)} ,\,
  P_{ggg  \leftarrow g}^{(0)},\,
  P_{gq\bar{q}  \leftarrow g}^{(0)},\,
  P_{\gamma q \bar{q} \leftarrow g}^{(0)}.
\label{eq:tcol_unpol_amplitudes}
\end{align}
The analytic expressions for the unpolarized triple-collinear splitting functions~\cite{Campbell:1997hg} are recalled in (\ref{eq:tcolfirst_unpol}\hspace{1pt}--\hspace{1pt}\ref{eq:tcollast_unpol}) in the Appendix.

From these  colour-ordered splitting functions ${P}_{(\Delta)ijk \leftarrow a}$,
 the $ \mathbf{P}_{(\Delta)ijk \leftarrow a}$ can be assembled as 
 follows~\cite{Catani:1998nv}: 
 \begin{alignat}{2}
  &\mathbf{P}_{qgg  \leftarrow q}^{(0)} &&= \frac{1}{2} \,C_FC_A \, \left(P_{qg_2g_3  \leftarrow q}^{(0)} 
  + P_{qg_3g_2  \leftarrow q}^{(0)} - P_{q\gamma \gamma  \leftarrow q}^{(0)}
  \right) + C_F^2 \,
  P_{q\gamma \gamma  \leftarrow q}^{(0)}\,, \\
  &\mathbf{P}_{q\bar{q}^\prime q^\prime  \leftarrow q}^{(0)} &&= C_F T_R \,
  P_{q\bar{q}^\prime q^\prime  \leftarrow q}^{(0)} \,, \\
  &\mathbf{P}_{q\bar{q}q  \leftarrow q}^{(0)} &&= C_F \bigg(C_F - \frac{1}{2} C_A \bigg)\, 
   {P}_{q\bar{q}q  \leftarrow q}^{(0)} \,, \\
  &\mathbf{P}_{gq\bar{q}  \leftarrow g}^{(0)} &&= \frac{1}{2}\, C_AT_R \,
    \left(P_{gq\bar{q}  \leftarrow g}^{(0)} + P_{g\bar{q}q  \leftarrow g}^{(0)} -  P_{\gamma q \bar{q} \leftarrow g}^{(0)}
     \right) + C_FT_R\, P_{\gamma q \bar{q} \leftarrow g}^{(0)}      \,, \\
  &\mathbf{P}_{ggg  \leftarrow g}^{(0)} &&= \frac{1}{4}\, C_A^2 \, \left( 
       P_{g_ig_jg_k  \leftarrow g}^{(0)} + (5~\mbox{permutations of~} i,j,k) \, \right) \,.
 \end{alignat}

The results for the triple collinear splitting functions 
are expressed in the form of~\cite{Braun-White:2022rtg}, which makes iterated single-collinear limits inside the triple-collinear splitting functions manifest.
Iterated single-collinear limits are captured in the coefficients of $1/(s_{ij}s_{ijk})$, $1/(s_{ik}s_{ijk})$, and $1/(s_{jk}s_{ijk})$.
In these coefficients we recover the appropriate single-collinear splitting functions with iterated arguments, which serves as a further check.
The remaining kinematic structure is purely triple-collinear.
For this purpose the structures 
\begin{align}
\mathrm{Tr}(ijkl)=z_i s_{jk} - z_j s_{ik} + z_k s_{ij}
\label{eq:def_Trijkl}
\end{align}
and
\begin{align}
W_{ij} = (x_i s_{jk} - x_j s_{ik})^2 - \frac{2}{1-\varepsilon} \frac{x_i x_j x_k}{1-x_k}s_{ij}s_{ijk}
\label{eq:def_Wij}
\end{align}
are introduced in the triple-collinear splitting functions.

Any particle in the triple-collinear splitting functions is allowed to be polarized.
From~\eqref{eq:tcol_unpol_amplitudes} one obtains 16 independent polarized triple-collinear splitting functions,
\begin{align*}
  & P_{qgg  \leftarrow q}^{(0)} && \longrightarrow 
    && P_{\Delta q \myspace g \myspace g \leftarrow q}^{(0)}\,,~
       P_{\Delta g \myspace g \myspace q \leftarrow q}^{(0)} \,,~
       P_{g\myspace \Delta g \myspace q \leftarrow q}^{(0)} \,,\\
  & P_{q\gamma\gamma\leftarrow q}^{(0)} && \longrightarrow
    && P_{\Delta q\myspace \gamma \myspace \gamma \leftarrow q}^{(0)} \,,~
       P_{\Delta \gamma\myspace \gamma \myspace q \leftarrow q}^{(0)} \,,\\
  & P_{q \qp \qbp \leftarrow q}^{(0)} && \longrightarrow
    && P_{\Delta q \myspace \qp \myspace \qbp \leftarrow q}^{(0)} \,,~
       P_{\Delta \qp \myspace \qbp \myspace q \leftarrow q}^{(0)} \,,\\
  & P_{q\qb q\leftarrow q}^{(0)} && \longrightarrow
    && P_{\Delta q \myspace \qb \myspace q \leftarrow q}^{(0)}  \,,~
       P_{\Delta \qb \myspace q \myspace q \leftarrow q}^{(0)} \,,\\
  & P_{ggg\leftarrow g}^{(0)} && \longrightarrow
    && P_{\Delta g\myspace g \myspace g \leftarrow g}^{(0)} \,,~
       P_{g\myspace \Delta g \myspace g \leftarrow g}^{(0)} \,, \\
  & P_{gq\qb \leftarrow g} && \longrightarrow
    && P_{\Delta g \myspace q \myspace \qb \leftarrow g}^{(0)} \,,~
       P_{g \myspace \Delta q \myspace \qb \leftarrow g}^{(0)} \,,~
       P_{\Delta q \myspace \qb \myspace g \leftarrow g}^{(0)} \,,\\
  & P_{\gamma q \qb \leftarrow g}^{(0)} && \longrightarrow
    && P_{\Delta \gamma \myspace q \myspace \qb \leftarrow g}^{(0)} \,,~
       P_{\Delta q \myspace \qb \myspace \gamma \leftarrow g}^{(0)} \,, \numberthis
\end{align*}
where the order of the final state partons reflects the color ordering where applicable.
In this sense, in $P_{\Delta g \myspace g \myspace q \leftarrow q}^{(0)}$, the polarized gluon is adjacent to the unpolarized gluon but not to the quark, whereas in $P_{g \myspace \Delta g \myspace q \leftarrow q}^{(0)}$ it is adjacent to both.
Similarly, in $P_{g \myspace \Delta g \myspace g \leftarrow g}^{(0)}$, the polarized gluon is color-adjacent to the other two collinear gluons, but in $P_{\Delta g \myspace g \myspace g \leftarrow g}^{(0)}$ it is only adjacent to the radiated gluon in the middle of the cluster.
And in $P_{\Delta q \myspace \qb \myspace g \leftarrow g}^{(0)}$ the polarized quark is not adjacent to the gluon, but in $P_{g \myspace \Delta q \myspace \qb \leftarrow g}^{(0)}$ it is.

As described in 
Section~\ref{sec:comp}, the polarized triple-collinear splitting functions are extracted from the triple-collinear limits of the photon and 
neutralino decay tree-level four-parton decay matrix elements and cross-validated on the graviton 
decay matrix elements and on five-parton photon decay matrix elements.

Denoting the first, second and third particle in the subscript by $(1,2,3)$, the 16 polarized triple-collinear 
splitting functions read:
\begin{align}
\label{eq:tcolfirst}
 \begin{autobreak}
  \input{tcol_limits/tcol_parse_latex_repl__P_Dq_g_g-q__nab.out}
 \end{autobreak} \\
 \begin{autobreak}
\input{tcol_limits/tcol_parse_latex_repl__P_Dg_g_q-q__nab.out}
 \end{autobreak} \\
 \begin{autobreak}
\input{tcol_limits/tcol_parse_latex_repl__P_g_Dg_q-q__nab.out}
 \end{autobreak} \\
 \begin{autobreak}
\input{tcol_limits/tcol_parse_latex_repl__P_Dq_g_g-q__ab.out}
 \end{autobreak} \\
 \begin{autobreak}
\input{tcol_limits/tcol_parse_latex_repl__P_Dg_g_q-q__ab.out}
 \end{autobreak} \\
 \begin{autobreak}
\input{tcol_limits/tcol_parse_latex_repl__P_Dq_qp_qbp-q.out}
 \end{autobreak} \\
 \begin{autobreak}
\input{tcol_limits/tcol_parse_latex_repl__P_Dqp_qbp_q-q.out}
 \end{autobreak} \\
 \begin{autobreak}
\input{tcol_limits/tcol_parse_latex_repl__P_Dq_qb_q-q__id.out}
 \end{autobreak} \\
 \begin{autobreak}
\input{tcol_limits/tcol_parse_latex_repl__P_Dqb_q_q-q__id.out}
 \end{autobreak} \\
 \begin{autobreak}
\input{tcol_limits/tcol_parse_latex_repl__P_Dg_g_g-g__nab.out}
 \end{autobreak} \\
 \begin{autobreak}
\input{tcol_limits/tcol_parse_latex_repl__P_g_Dg_g-g__nab.out}
 \end{autobreak} \\
 \begin{autobreak}
\input{tcol_limits/tcol_parse_latex_repl__P_Dg_q_qb-g__nab.out}
 \end{autobreak} \\
  \begin{autobreak}
\input{tcol_limits/tcol_parse_latex_repl__P_g_Dq_qb-g__nab.out}
 \end{autobreak} \\
  \begin{autobreak}
\input{tcol_limits/tcol_parse_latex_repl__P_Dq_qb_g-g__nab.out}
 \end{autobreak} \\
 \begin{autobreak}
\input{tcol_limits/tcol_parse_latex_repl__P_Dg_q_qb-g__ab.out}
 \end{autobreak} \\
 \begin{autobreak}
\input{tcol_limits/tcol_parse_latex_repl__P_Dq_qb_g-g__ab.out}
 \end{autobreak}
 \label{eq:tcollast}
\end{align}

The decomposition into the form of~\cite{Braun-White:2022rtg}
and its 
underlying assumptions on iterated limits 
have so far only been verified for the unpolarized triple-collinear splitting functions. We find that 
the polarized triple-collinear splitting functions admit the same decomposition
into invariant structures. 
To arrive at this form, we use the partial fraction decomposition algorithm inside a Gröbner basis with a monomial ordering relation from \texttt{MultivariateApart}~\cite{Heller:2021qkz}, which provides a linear basis of 19 pole structures.
We have verified that the 19 structures are equivalent to the 19 pole structures found in (4.1) of~\cite{Braun-White:2022rtg}.
We also employ an identical partial fraction decomposition to map the 13 invariant structures of (4.2) in~\cite{Braun-White:2022rtg} onto the basis of 19 pole structures and identify a linear map from the physically overcomplete basis onto the 13 invariant structures.
By identifying unique structures arising in the partial fraction in terms of 13 invariant structures, we are able to map each of the polarized triple-collinear splitting functions expressed in terms of the 19 pole structures onto the 13 invariant structures without retaining leftover terms which do not fit into the 13 invariant structures.

We also calculate the initial-final limits of the antenna functions.
For DIS we restrict ourselves to the case where the polarized parton is crossed into the initial state if it is present inside the collinear cluster. As outlined in Section~\ref{sec:kin} above, the 
splitting functions in initial-final kinematics for parton $i$ in the initial state can be obtained 
from (\ref{eq:tcolfirst}\hspace{1pt}--\hspace{1pt}\ref{eq:tcollast}) 
by 
substituting $z_i = 1-\sum_{r\neq i} z_r$, followed by the replacement $z_r=x_r/(1-\sum_{s\neq i} x_s)$, 
which we confirmed by explicit calculation of the triple collinear limits of four-particle 
matrix elements in DIS kinematics.

\section{Conclusions and outlook}
\label{sec:conc}

NNLO QCD calculations for benchmark observables in collisions of polarized particles will require 
a subtraction scheme to handle infrared-singular real-radiation contributions in a process-independent manner. 
To construct such a subtraction scheme, the behaviour of helicity-dependent matrix elements in their
  infrared-unresolved limits  must be determined. Polarization is relevant only for uniquely identified 
  polarized partons, such that the only unresolved configurations of interest are collinear splittings involving 
  one polarized and up to two unpolarized partons. 
  
  In these limits, matrix elements factorize into the product of a splitting function and a reduced matrix element 
where the collinear cluster is replaced by a particle of combined momentum and quantum numbers. At NNLO, 
single-collinear splitting functions enter up to one loop and triple-collinear splitting functions at tree-level. 
By computing the single and triple collinear limits of matrix elements for the decay of an external
 current into three or four partons, we re-derived the unpolarized NNLO splitting functions 
and newly obtain the polarized NNLO splitting functions. Our results 
are expressed in a colour-ordered from and imply an average over the transverse orientation of 
the collinear cluster. We employ the Larin scheme for helicity projectors of partons and of the decay current 
throughout and provide expressions for the unrenormalized splitting functions. 

The newly derived polarized splitting functions allow to construct infrared subtraction terms 
that extract the triple collinear behaviour of polarized tree-level double-real radiation matrix elements and 
the single collinear behaviour of polarized one-loop single-real radiation matrix elements. This construction could 
for example follow the designer antenna approach~\cite{Braun-White:2023sgd,Braun-White:2023zwd,Fox:2023bma,Fox:2024bfp}, where different configurations
 of unresolved emissions off hard radiator partons 
are iteratively assembled into so-called antenna functions. The decomposition of triple collinear splitting functions 
into iterated and genuine contributions~\cite{Braun-White:2022rtg} already prepares for such a construction. 
The precise form of the analytical integration of the resulting 
subtraction terms is different in each subtraction method. The splitting functions derived here correspond to the 
universal residues that are associated with the unresolved collinear behaviour. They will thus fully determine 
the infrared pole structure of the integrated subtraction terms in any subtraction method and enable
the cancellation of infrared singularities between real radiation and virtual 
corrections to a given process.

\acknowledgments
We would like to thank Oscar Braun-White, Prasanna K.\ Dhani, Nigel Glover, and Kay Sch\"onwald for discussions. 
This work has received funding from the Swiss National Science Foundation (SNF)
under contracts  200020-204200 and 240015 and from the European Research Council (ERC) under
the European Union's Horizon 2020 research and innovation programme grant
agreement 101019620 (ERC Advanced Grant TOPUP).

\appendix
\section{Unpolarized triple-collinear splitting functions}

The unpolarized colour-ordered and angular-averaged
 triple-collinear splitting functions~\cite{Campbell:1997hg} can be expressed in a form that separates their 
 iterated single collinear and genuine triple collinear contributions. They are then 
  given by~\cite{Braun-White:2022rtg}:
\begin{align}
\label{eq:tcolfirst_unpol}
 \begin{autobreak}
  \input{tcol_limits/tcol_parse_latex_repl__P_q_g_g-q__nab.out}
 \end{autobreak} \\
 \begin{autobreak}
\input{tcol_limits/tcol_parse_latex_repl__P_q_g_g-q__ab.out}
 \end{autobreak} \\
 \begin{autobreak}
\input{tcol_limits/tcol_parse_latex_repl__P_q_qp_qbp-q.out}
 \end{autobreak} \\
 \begin{autobreak}
\input{tcol_limits/tcol_parse_latex_repl__P_q_qb_q-q__id.out}
 \end{autobreak} \\
 \begin{autobreak}
\input{tcol_limits/tcol_parse_latex_repl__P_g_g_g-g__nab.out}
 \end{autobreak} \\
 \begin{autobreak}
\input{tcol_limits/tcol_parse_latex_repl__P_q_qb_g-g__nab.out}
 \end{autobreak} \\
 \begin{autobreak}
\input{tcol_limits/tcol_parse_latex_repl__P_q_qb_g-g__ab.out}
 \end{autobreak}
 \label{eq:tcollast_unpol}
\end{align}
The symbols $\operatorname{Tr}(ijkl)$ and $W_{ij}$ are defined as in~\eqref{eq:def_Trijkl} and \eqref{eq:def_Wij}.

\bibliographystyle{JHEP}
\bibliography{pol_limits}

\end{document}